# New approach for fabrication germanene with Dirac electrons preserved: A first principle study

Ping Li, Juexian Cao and Zhi-Xin Guo[a)]

*Department of Physics, Xiangtan University, Xiangtan, Hunan, 411105, China*

## Abstract

How to obtain germanene with Dirac electrons preserved is still an open challenge. Here we report a sandwich-dehydrogenation approach, i.e., to fabricate germanene through dehydrogenating germanane in a sandwiched structure. The dehydrogenation can spontaneously occur for the sandwiched structure, which overcomes the problem of amorphization in the heating dehydrogenation approach. The obtained germanene preserve the Dirac electronic properties very well. Moreover, the Fermi velocity of germanene can be efficiently manipulated through controlling the interlayer spacing between germanane and the sandwiching surfaces. Our results indicate a guideline for fabrication of prefect two-dimensional materials.

Corresponding authors:
zxguo08@hotmail.com

# 1. Introduction

In recent years, tow-dimensional (2D) materials have attracted much attention due to their unique properties. A representative example is graphene, in which $sp^2$ hybridized electrons (σ electrons) form a honeycomb structure and the remaining $P_z$ (π) electrons follow the massless Dirac (Weyl) equation [1-4]. The energy bands show the linear dispersion (Dirac cone) at the Fermi level ($E_F$) at particular symmetry points, *K* and *K'*, in Brillouin zone (BZ) [5]. This gives rise to unique phenomena such as the anomalous quantum Hall effect and unexpected magnetic ordering [6], which make graphene a promising material utilized for the next-generation nanodevices [7]. Yet, the nature of tiny topological band gap (~1meV) and the issue of integrating graphene in current Si-, Ge-based nano/micro-electronics devices restrict its applications [8].

Recently, a 2D sheet consisting of Ge (germanene) has been both theoretically predicted and experimentally synthesized [9-13]. It has been clarified that the freestanding germanene presents a honeycomb structure with the buckling and exhibits the graphene-like band structure around $E_F$ supporting the existence of Dirac electrons. The larger topological band gap (~34meV) induced by strong spin-orbit interaction in Ge and the strong quantum spin-Hall phase is another intriguing factor in science viewpoints [14]. Combined with the good compatibility with modern semiconductor technology, germanene has great application prospects in the next-generation nanodevices [15,16].

By far germanene is obtained only via growing on the metallic substrates,

which have strong interaction with germanene [11-13]. Recently, Bianco *et al.* synthesized the millimeter-scale crystals of multi-layered germanane (H-Ge) [17]. But they failed to obtain germanene because it needs to heat up to 200℃ to 250℃ for the dehydrogenation whereas the amorphization occurs above 75℃ [17]. Therefore, how to obtain germanene with Dirac electrons preserved is still an open challenge.

In this work, based on the first-principles calculations we propose a new approach to fabricating germanene, i.e., dehydrogenating H-Ge in a sandwiched structure. We find that the dehydrogenation can spontaneously occur for H-Ge sandwiched structure. The band structure calculations demonstrated that the obtained germanene exhibits Dirac electronic properties. This approach overcomes the problem of amorphization in the heating dehydrogenation method.

## 2. Methods

Density functional theory (DFT) calculations were carried out by using the projector augmented wave method and the generalized gradient approximation (GGA), as implemented in the Vienna *ab initio* simulation package (VASP) [18-20]. The exchange-correlation functional, vdW-DF, being capable of treating the van der Waals (vdW) force was adopted (optB86b-vdW functional) [21-23]. The projector-augmented-wave pseudopotential with an energy cutoff of 400 eV for the plane-wave basis sets was adopted, and Gamma-centered 15×15×1 Monkhorst-Pack k-point meshes were used for slab calculations [24,25]. The convergence criteria were set to $1\times10^{-5}$ eV for total energy and 0.005 eV/Å for Hellman-Feynman forces respectively.

Our system was modeled as a monolayer of H-Ge sandwiched by two surfaces as shown in Fig. 1. For comparison, our considering surfaces involve hydrogen-saturated Si(111), Ge(111), Ga-/As-terminated GaAs(111) (denoted as Ga-GaAs, As-GaAs) surface, graphene, single h-BN sheet and also metallic Au(111) surface. The surfaces are simulated by a repeated slab model consisting of four or five atomic layers cleaved from their bulk. The lattice constants of all surfaces are fixed to their experimental values. Taking the lattice mismatch into account, we choose 1×1 H-Ge on 1×1 Ge(111) and 1×1 GaAs(111) while 2×2 superperiodicities of H-Ge on 3×3 Au(111), 3×3 graphene and 3×3 h-BN respectively. A vacuum region of 20 Å was applied along the z-direction, which is large enough to make the interaction between neighboring images negligible.

## 3. Result and discussion

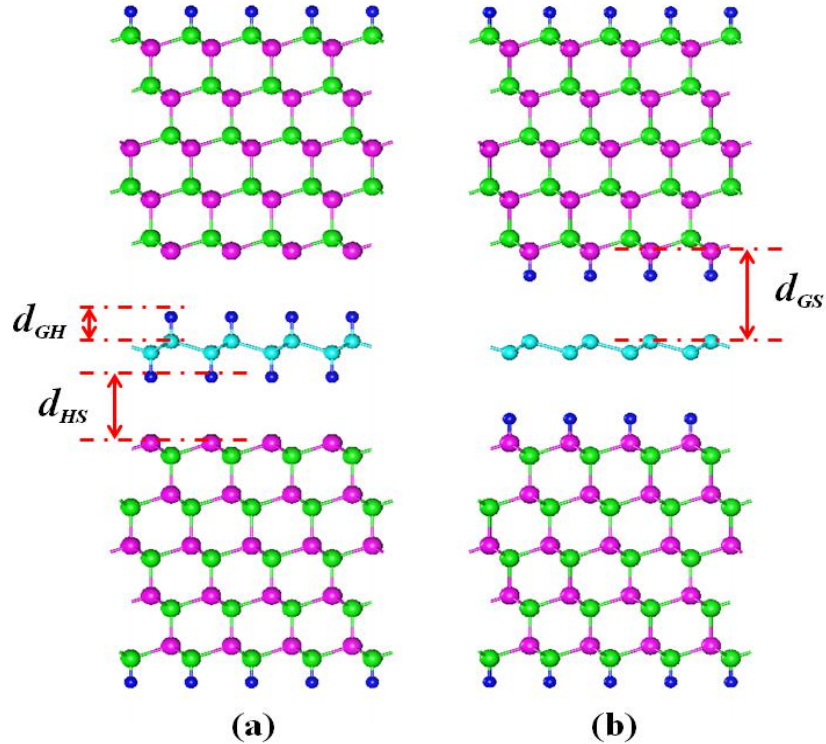

**Fig. 1.** (Color online) Atomic configuration of H-Ge sandwiched by hydrogen-saturated Ga-terminated GaAs(111). The green, red, light blue and dark blue bass represent As, Ga, Ge and H atoms, respectively. $d_{GH}$, $d_{HS}$ and $d_{GS}$ indicate the bond length of H-Ge in hydrogenated germanene, the distance between the hydrogen in hydrogenated germanene and the outmost Ga in GaAs(111) surface, the average distance between hydrogenated germanene and GaAs(111) surface.

Our calculations start with the possibility of dehydrogenation for H-Ge which is sandwiched by two substrates as shown in Fig. 1. The distance $d_{GS}$ between the substrate and the H-Ge is initially set to be 4 Å. For convenience, we take the Ga-GaAs(111) surface as an example as shown in Fig. 1. After fully relaxation, we found that the hydrogen atoms automatically moves away from the H-Ge to the substrates and tightly bonded to Ga in GaAs(111) surface, leaving the monolayer of Ge alone in the middle as shown in Fig. 1(b). Our calculation clearly demonstrates that the dehydrogenation spontaneously occurs for the H-Ge sandwiched by the Ga-GaAs(111) surface. The dehydrogenation is further confirmed for the H-Ge sandwiched by As-GaAs(111), Si(111) and also Ge(111) surface. But it is failed for graphene, h-BN and Au(111) surfaces due to the fact that the hydrogen atoms remain tightly bonded to the germanene. The structural parameters for the dehydrogenated germanene are shown in Table I. One can find that the buckling Δd (0.68-0.70 Å) and the bond lengths $l$ (2.45-2.47 Å) are very similar to those of freestanding germanane (Δd=0.68 Å, $l$=2.45 Å)

**Table I.** The calculated geometry parameters for the obtained germanene after dehydrogenation, as well as the bader change, the surface work function Φ, and the dehydrogenation energy $E_D$. The *a, l,* and Δ*d* represent the lattice constant, Ge-Ge bond length, and the buckling of obtained germanene, respectively. $d_{GS}$ represents the interlayer spacing between the sandwiched Ge layer and the surface.

|  | a(Å) | $d_{GS}$(Å) | l(Å) | Δd(Å) | bader | Φ(eV) | $E_D$(eV) |
|---|---|---|---|---|---|---|---|
| **Si(111)** | 4.066 | 4.10 | 2.460 | 0.6933 | 0.0586 | 5.064 | 0.815 |
| **Ge(111)** | 4.066 | 3.93 | 2.467 | 0.6854 | -0.0315 | 4.668 | 0.414 |
| **Ga-GaAs(111)** | 4.083 | 3.81 | 2.454 | 0.6821 | -0.4630 | 4.494 | 0.321 |
| **As-GaAs(111)** | 4.083 | 3.79 | 2.473 | 0.6986 | 0.4661 | 5.107 | 0.396 |

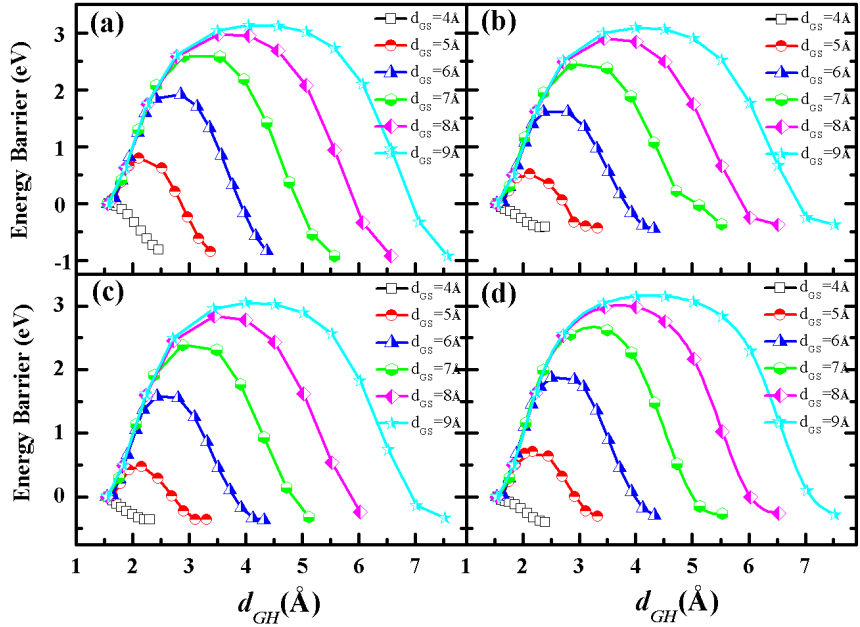

**Fig. 2.** (Color online) Calculated reaction energy barrier profile for the dehydrogenation H-Ge sandwiched between surfaces of Si(111) (a), Ge(111) (b), Ga-GaAs(111) (c), and As-GaAs(111) (d) as a function of interlayer spacing between the Ge layer and the surface $d_{GS}$.

To interpret the validity of dehydrogenation, the hydrogen dissociation energy ($E_D$) is additionally calculated given in Table I. The hydrogen dissociation energy $E_D$ is defined as $E_D = \frac{1}{N}(E_{sub} + E_{H-Ge} - E_{tot})$, where $E_{sub}$, $E_{H-Ge}$ and $E_{tot}$ are the total energies of the two substrates, hydrogenated germanene and sandwiched structure shown in Fig. 1, respectively. $N$ is the number of hydrogen atoms in hydrogenated germanene. As shown in Table I, $E_D$ is in the range of 0.29-0.84 eV per H atom for the Si(111), Ge(111) Ga-GaAs(111) and As-GaAs(111) sandwiched structures. The large positive value of $E_D$ indicates that the reaction of dehydrogenation is exothermic reaction and energetically favorable. To further understand to the process of dehydrogenation for H-Ge in detail, the reaction path and reaction barrier are calculated as shown in Fig. 2. The reaction energy barrier and the reaction path are quite similar for Si(111), Ge(111), Ga-GaAs(111) and As-GaAs(111) surface. Especially, the energy barrier is almost zero which indicates that the dehydrogenation process spontaneously occurs for the distance between H-Ge and surface $d_{GS}$ smaller than 4 Å. The energy barrier increases with the increase of $d_{GS}$. The $d_{GS}$ dependence of dehydrogenation barrier for the Si(111), Ge(111), Ga-GaAs(111) and As-GaAs(111) surfaces have been summarized in Fig. 3. Obviously, the energy barriers remain the same value of 3.0 eV when the $d_{GS}>9$ Å. Because H-Ge can be thermally stable up to 75℃ [17], the dehydrogenation is expected to be realized with the effect of temperature as the energy barrier is smaller than 1 eV/H for $d_{GS}<5$ Å.

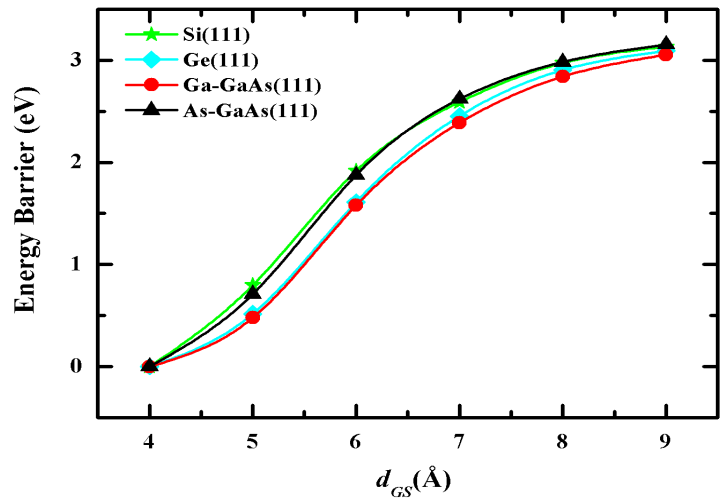

**Fig. 3.** (Color online) Dehydrogenation barrier of H-Ge sandwiched between surfaces of Si(111) (green stars), Ge(111) (blue rhombuses), Ga-GaAs(111) (red circles), and As-GaAs(111) (black triangles) as a function of

$d_{GS}$, respectively.

As shown in Fig. 3, the energy barriers for the Si(111) and As-GaAs(111) are larger than that for Ge(111) and Ga-GaAs(111) in the entire $d_{GS}$ range. The results obviously demonstrate that the dehydrogenation barriers depend on the surface material. To reveal the underlying mechanism of surface material effect on dehydrogenation, we have calculated the bader charge and work function Φ for the surface material listed in Table I [26,27]. In general, the dehydrogenation depends on the ability of capturing hydrogen by surface. It means that the ability of dehydrogenation corresponds to the ability of obtaining electrons. It is known that the larger bader charge corresponds to the weaker ability of obtaining electron. And also, the larger work function represents the weaker losing-electron capacity and thus the poorer obtaining-electron ability. As shown in Table I, the bader charge and work function in the Si(111) and As-GaAs(111) surfaces are indeed larger than those in the Ge (111) and Ga-GaAs(111) surfaces. This explains why the dehydrogenation barriers for former two surfaces are larger than that for the latter two.

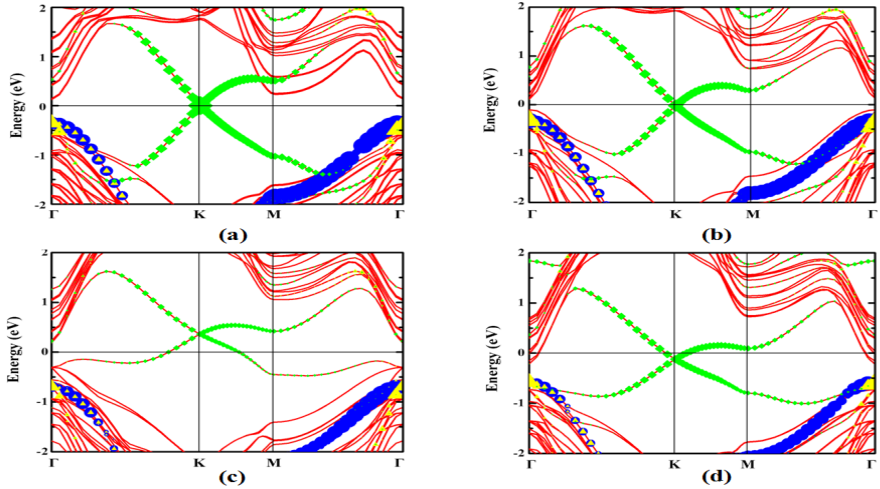

**Fig. 4.** (Color online) Calculated energy bands of the stable structures for the 1×1 germanene layer sandwiched between Si(111) (a), Ge(111) (b), Ga-GaAs(111) (c), and As-GaAs(111) (d) surfaces after

dehydrogenation. The orbital projections for $4P_z$ (green rhombus), $4P_x$ (blue circle), and $4P_y$ (yellow triangle) on energy bands of germanene are also indicated. The linear energy bands near Fermi level are from the $4P_z$ orbital of germanene, showing that the Dirac electrons are well preserved.

We have further calculated the electronic band structures for H-Ge sandwiched by the four types of surface after dehydrogenation. As shown in Fig. 4, there are linear energy bands at $K$ point in BZ around $E_F$. Through the analysis of orbital projections on energy bands, it is found that the linear bands have almost entirely Ge $4P_z$ orbital character (green circles in Fig. 4). The contribution of other states of Ge and the surface elements is extremely weak. This clearly shows that the Dirac electrons have been well preserved after the dehydrogenation of H-Ge by the sandwiched structure.

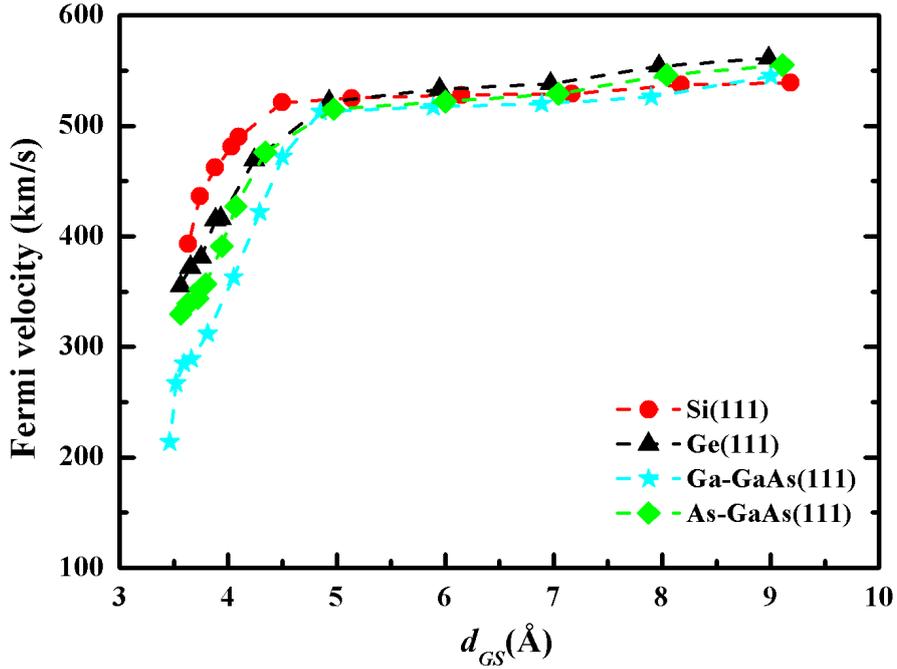

**Fig. 5.** (Color online) Calculated Fermi velocity ($V_F$) of Dirac electrons as function of $d_{GS}$ for germanene sandwiched between surfaces of Si(111) (red circles), Ge(111) (black triangles), Ga-GaAs(111) (blue stars) and As-GaAs(111) (green rhombuses), respectively.

Another interesting issue to be clarified is the influence of surfaces on the Fermi velocity ($V_F$) of dehydrogenated germanene [28]. With variation of $d_{GS}$, we have systematically calculated $V_F$ of Dirac electrons for the germanene sandwiched between the surfaces of Si(111), Ge(111), Ga-GaAs(111) and As-GaAs(111). As shown in Fig. 5, for all the four sandwiching surfaces $V_F$ monotonically increases with the increase of the $d_{GS}$, and approaches to a constant value 550 km/s, i.e., the $V_F$ of freestanding germanene. This shows the nature of suppressing $V_F$ of germanene in the sandwich structures. It is noted that there a inflection for $V_F$ around $d_{GS}$=4.5 Å, below which the variation of $V_F$ becomes drastic due to the strong interlayer interactions between germanene and the sandwiching surfaces induced by the small $d_{GS}$. The above results lighten an efficient way on manipulating $V_F$ of germanene.

Finally, we propose an experimental procedure to realize our sandwich-dehydrogenation approach. Firstly, a monolayer of H-Ge is obtained by the mechanical exfoliation from the bulk H-Ge following the way of obtaining graphene from graphite. Then, the obtained H-Ge is placed on the substrate surfaces such as Si(111), Ge(111), GaAs(111), etc.. Finally, antoher thin films of Si(111), Ge(111), GaAs(111), etc., are placed/deposited on the H-Ge to form the sandwiched structure [29-32]. From our calculations, the vdW-equilibrium distance between gremanene and sandwiched surfaces is around 4 Å. This means that the dehydrogenation spontaneously occurs once the sandwich structure is formed in usual case. In the particular cases where $d_{GS}$ is larger than 5 Å, the dehydrogenation can be realized by decreasing $d_{GS}$ in use of external pressure or electric field [33], both of which can efficiently decrease the energy barrier for dehydrogenation.

## 4.Conclusion

In summary, by using the first-principles calculations we proposed a new approach for germanene fabrication via dehydrogenating H-Ge in use of the surface effect. We clearly demonstrated that the prefect germanene can be obtained by the dehydrogenation process for H-Ge sandwiched by surfaces. The dehydrogenation barrier significantly decreases with the decrease of the interlayer distance $d_{GS}$. The new proposed approach for germanene fabrication overcomes the problem of amorphization in the heating dehydrogenation approach. Most importantly, the Dirac electrons properties are preserved for the obtained germanene. Since the obtained germanene is naturally encapsulated by the sandwiched surfaces after dehydrogenation, our approach has the superiority of protecting germanene from environmental pollution, which makes it easier applied in the nanodevices.

## Acknowlegements

This work was supported by the National Natural Science Foundation of China (Grant No. 11204259, 11474245, 11374252), Natural Science Foundation of Hunan Province (No. 2015JJ6106), the Program for New Century Excellent Talents in University of Ministry of Education of China (Grant No.NCET-12-0722).

## Captions of figures and tables

**Fig. 1.** (Color online) Atomic configuration of H-Ge sandwiched by hydrogen-saturated Ga-terminated GaAs. The green, red, light blue and dark blue bass represent As, Ga, Ge and H atoms, respectively. $d_{GH}$, $d_{HS}$ and $d_{GS}$ indicate the bond length of H-Ge in hydrogenated germanane, the distance between the hydrogen in hydrogenated germanane and the outmost Ga in GaAs(111) surface, the average distance between hydrogenated germanane and GaAs(111) surface.

**Fig. 2.** (Color online) Calculated reaction energy barrier profile for the dehydrogenation H-Ge sandwiched between surfaces of Si(111) (a), Ge(111) (b), Ga-GaAs(111) (c), and As-GaAs(111) (d) as a function of interlayer spacing between the Ge layer and the surface $d_{GS}$.

**Fig. 3.** (Color online) Dehydrogenation barrier of H-Ge sandwiched between surfaces of Si(111) (green stars), Ge(111) (blue rhombuses), Ga-GaAs(111) (red circles), and As-GaAs(111) (black triangles) as a function of $d_{GS}$, respectively.

**Fig. 4.** (Color online) Calculated energy bands of the stable structures for the 1×1 germanene layer sandwiched between Si(111) (a), Ge(111) (b), Ga-GaAs(111) (c), and As-GaAs(111) (d) surfaces after dehydrogenation. The orbital projections for $4P_z$ (green rhombus), $4P_x$ (blue circle), and $4P_y$ (yellow triangle) on energy bands of germanene are also indicated. The linear energy bands near Fermi level are from the $4P_z$ orbital of germanene, showing that the Dirac electrons are well preserved.

**Fig. 5.** (Color online) Calculated Fermi velocity ($V_F$) of Dirac electrons as function of $d_{GS}$ for germanene sandwiched between surfaces of Si(111) (red circles), Ge(111) (black triangles), Ga-GaAs(111) (blue stars) and As-GaAs(111) (green rhombuses), respectively.

**Table I.** The calculated geometry parameters for the obtained germanene after dehydrogenation, as well as the bader change, the surface work function Φ, and the dehydrogenation energy $E_D$. The *a, l*, and $\Delta d$ represent the lattice constant, Ge-Ge bond length, and the buckling of obtained germanene, respectively. $d_{GS}$ represents the interlayer spacing between the sandwiched Ge layer and the surfaces.

# Figures and Tables

**Fig.1.** Li et.al

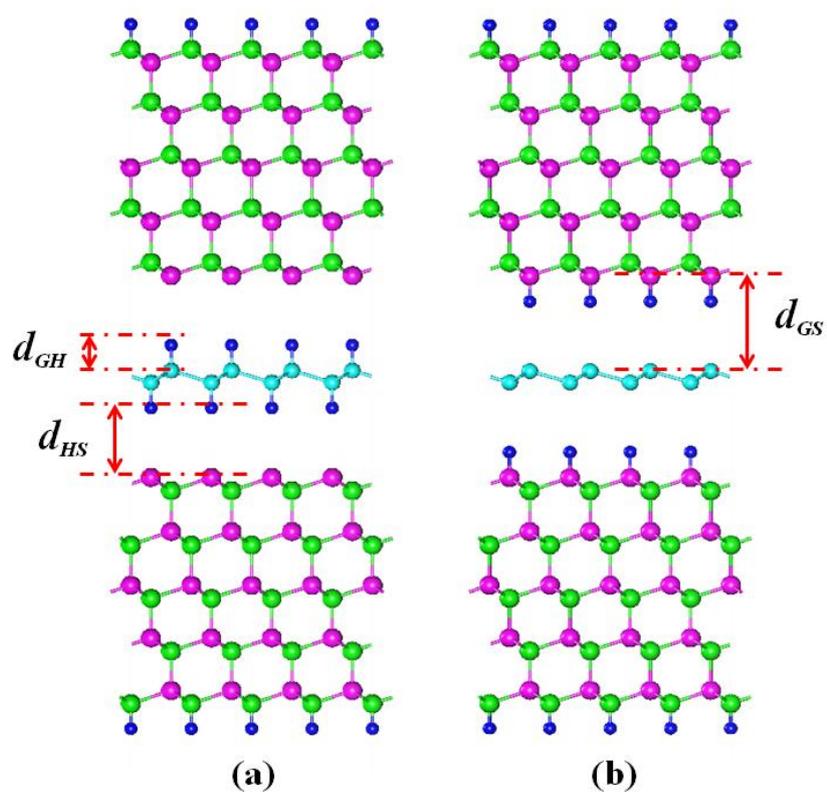

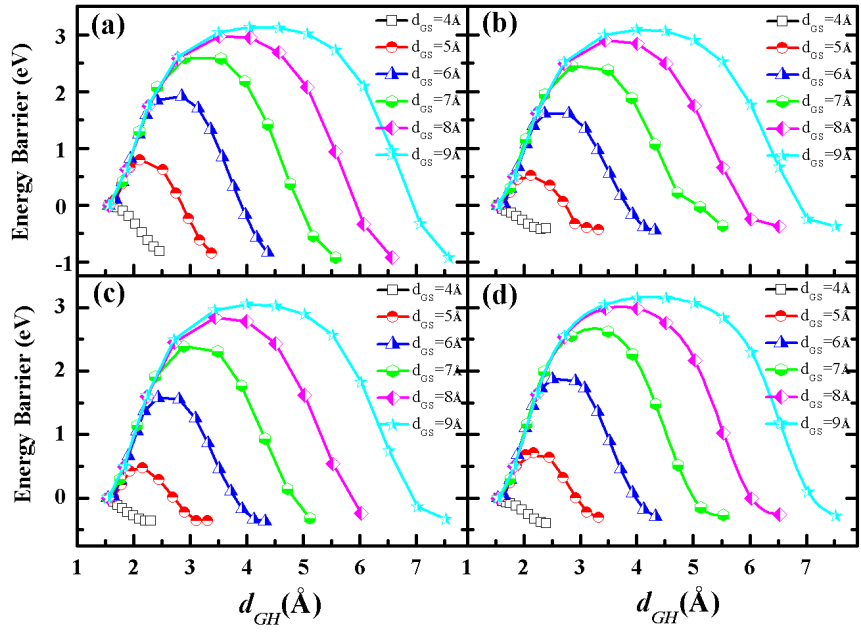

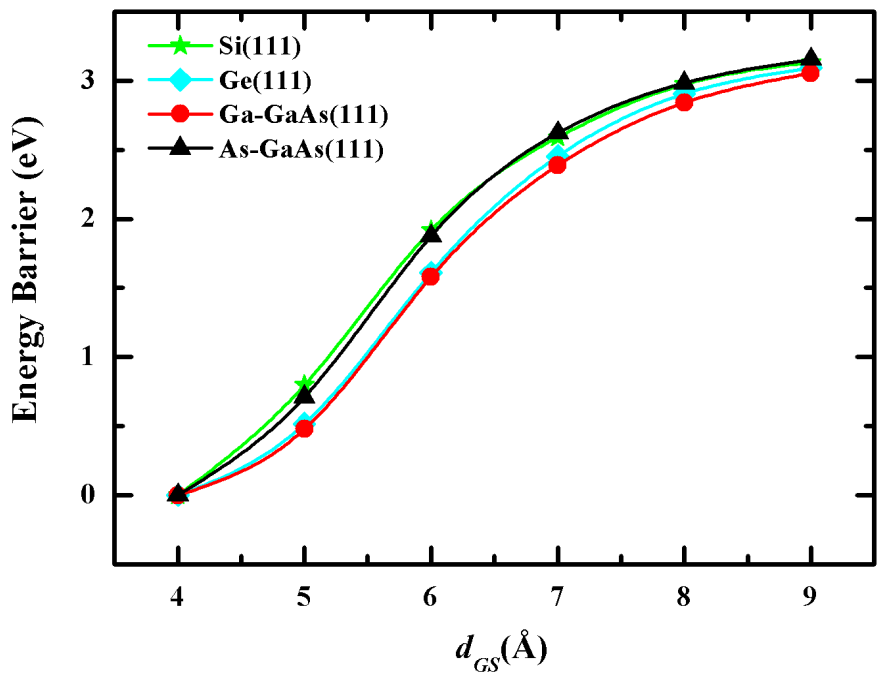

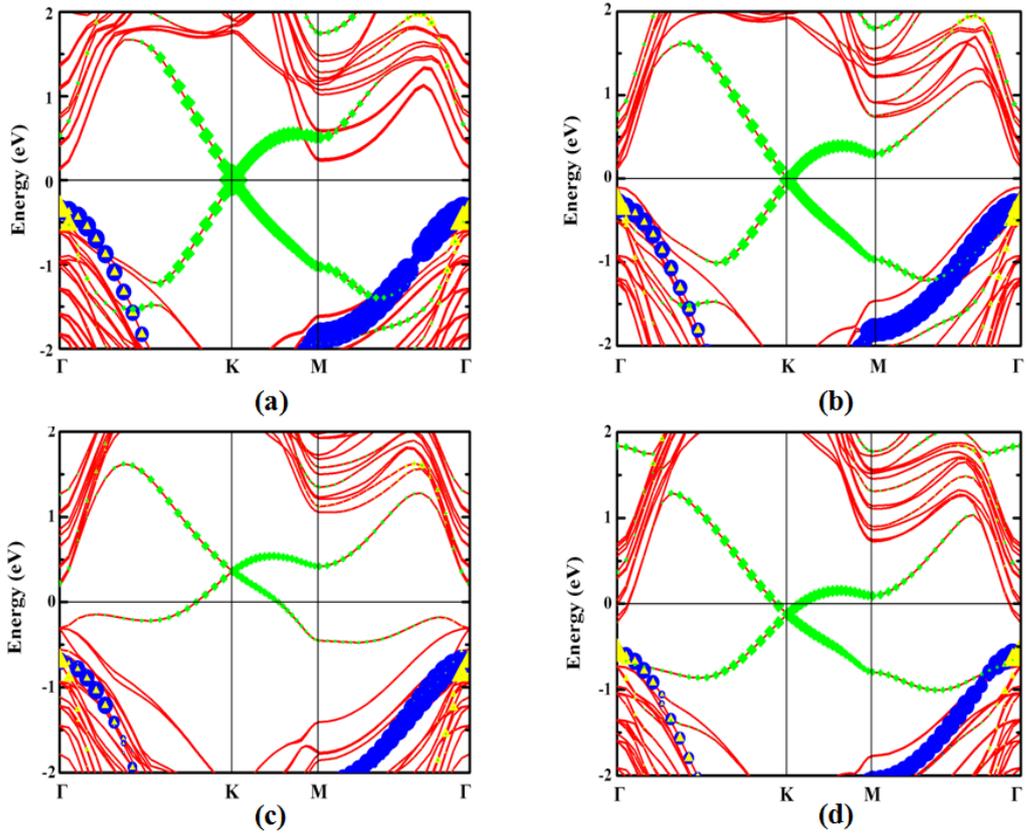

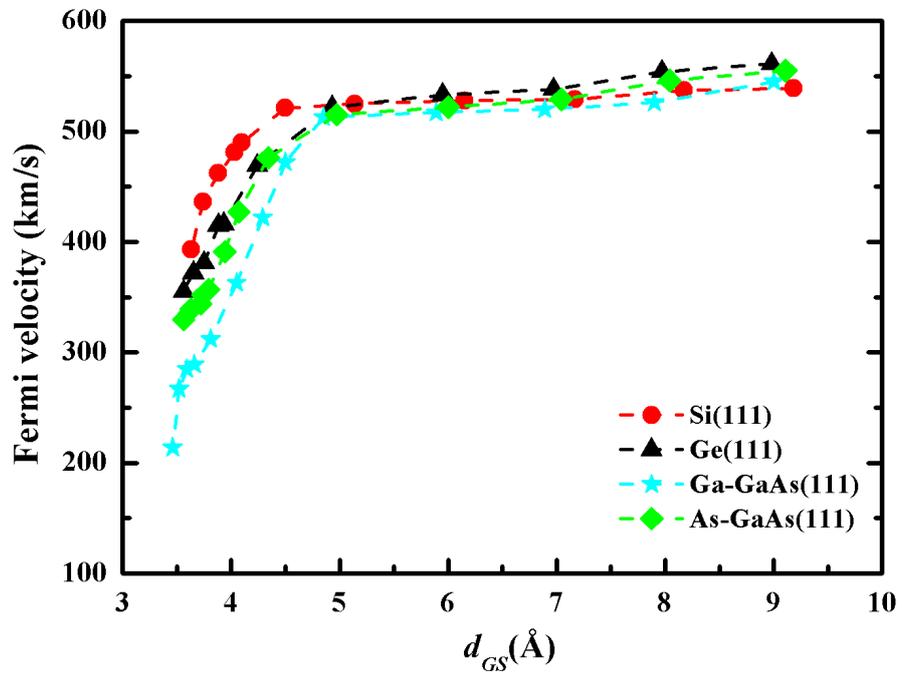

|  | a(Å) | $d_{GS}$(Å) | $l$(Å) | $\Delta d$(Å) | bader | $\Phi$(eV) | $E_D$(eV) |
|---|---|---|---|---|---|---|---|
| **Si(111)** | 4.066 | 4.10 | 2.460 | 0.6933 | 0.0586 | 5.064 | 0.815 |
| **Ge(111)** | 4.066 | 3.93 | 2.467 | 0.6854 | -0.0315 | 4.668 | 0.414 |
| **Ga-GaAs(111)** | 4.083 | 3.81 | 2.454 | 0.6821 | -0.4630 | 4.494 | 0.321 |
| **As-GaAs(111)** | 4.083 | 3.79 | 2.473 | 0.6986 | 0.4661 | 5.107 | 0.396 |